# Quantum vacuum effects for a massive Bosonic string in background B-field

Y Koohsarian[1] and A Shirzad[2,3]

1. Department of Physics, Ferdowsi University of Mashhad, Iran
2. Department of Physics, Isfahan University of Technology, Isfahan, Iran
3. School of Physics, Institute for Research in Fundamental Sciences (IPM), Tehran, Iran

E-mail: yo.koohsarian@stu-mail.um.ac.ir

**Abstract**
We study the Casimir effect for a Bosonic string extended between D-branes, and living in a flat space with an antisymmetric background B-field. We find the Casimir energy as a function of the B-field, and the mass-parameter of the string, and accordingly we obtain a B-dependence correction term to the ground-state mass of the string. We show that for sufficiently large B-field, the ground state of the string contains real (i.e. non-Tachyonic) particles.

## 1. Introduction

Imposing definite boundary conditions on a quantum field changes the spectrum of quantum states and leads in particular to changing the vacuum energy of the system. This change to the quantum vacuum, results in some observable quantum effects such as the well-known Casimir force [1]. As it is known, this force depends specifically on the geometrical features of the system, such as the specific boundary conditions imposed on the field.

A variety of theoretical and experimental models with different boundary conditions have been considered in the framework of Casimir effect (see e.g. [2, 3] and [22] as a review). The general procedure is to find Hamiltonian of the system as a combination of physical (mostly harmonic oscillation) modes, which have quantum ground states with nonzero energies. Then the vacuum state energy would be found as a summation over the zero-point energies of these physical oscillatory modes. The Casimir energy is obtained by subtracting the contribution of free (i.e. unbounded space) from the vacuum energy.

In connection with the string theory, the Casimir effect has been investigated from different perspectives [4-12]. In this paper the model of an open Bosonic string (with a nonzero mass-parameter) ending on D-branes, and having a background B-field, has been considered, with a background B-field, introduced initially in [13]. This model is a generalization of the zero mass-parameter case, which is a famous model in the context of the string theory, especially because of exhibiting noncommutative coordinates on the branes attached to the endpoints of the string [13-16]. Note that the open Bosonic string theory (with a 25-dimensional space) still has important applications, such as the famous problem of the Tachyonic modes [17-20]. In a previous paper [21], considering the boundary conditions as Dirac constraints and imposing them on Fourier expansions of the fields, we found the physical modes of the system as an infinite set of harmonic oscillators. This enabled us to write down canonical Hamiltonian as a summation over Hamiltonian of simple harmonic oscillators with definite frequencies. Hence, we can read out the vacuum energy as the summation over the zero-point energy of individual oscillators and regularize it to find out the Casimir energy of the string. We apply the well-known Abel-Plana formula for regularization of the vacuum energy to find the Casimir energy of the string. Then utilizing this Casimir energy, we find correction terms to the ground-state mass of the Bosonic string.

## 2. Casimir effect for the massive Bosonic string

First, we briefly review the approach given in [21], to find the Hamiltonian of the string. Suppose an even number of fields, $X^i$ among Bosonic fields $X^\mu$ living in a flat target space, are coupled to an antisymmetric external B-field. In the simplest case, the subspace of $X^i$'s is a two dimensional Euclidian space, and the constant B-field is exhibited by

$$B = \begin{pmatrix} 0 & \tilde{B} \\ -\tilde{B} & 0 \end{pmatrix}. \tag{1}$$

Thus, neglecting those components of $X^\mu$ which do not couple to B-field, the simplified Lagrangian is given as [21]

$$L = \frac{1}{2}\int_0^\ell d\sigma \, [\dot{X}^i \dot{X}_i - X'^i X'_i - m^2 X^i X_i + 2 B_{ij} \dot{X}^i X'_i], \tag{2}$$

where "dot" and "prime" represent differentiation with respect to $\sigma$ and $\tau$ respectively, $m$ is the mass-parameter of the Bosonic fields, and $\ell$ is the length-parameter of the string; this is the simplified version of the model given in [13]. In canonical formulation the Hamiltonian reads

$$H = \frac{1}{2}\int_0^\ell [(P_i - B_{ij} X'_j)^2 + X'^2 + m^2 X^2] d\sigma, \tag{3}$$

where $P_i = \dot{X}_i + B_{ij} X'_j$ are conjugate momentum operators. The equation of motion would be obtained as $(\partial_\tau^2 - \partial_\sigma^2 - m^2) X_i = 0$ with the boundary condition (BC) as $X'_i + B_{ij} \dot{X}_j = 0$ and, as shown in details in [21], the solutions are found as

$$\begin{aligned}X(\sigma,\tau) &= \frac{1}{\sqrt{\ell}}[a_0(\tau)\cosh k_0(\sigma-\ell/2) \\ &\quad - \frac{1}{k_0} M^{-1} B c_0(\tau)\sinh k_0(\sigma-\ell/2)] \\ &\quad + \sqrt{\frac{2}{\ell}}\sum_{n=1}^\infty [a_n(\tau)\cos\frac{n\pi}{\ell}\sigma - \frac{\ell}{n\pi} M^{-1} B c_n(\tau)\sin\frac{n\pi}{\ell}\sigma], \\ P(\sigma,\tau) &= \frac{1}{\sqrt{\ell}}[c_0(\tau)\cosh k_0(\sigma-\ell/2) \\ &\quad - \frac{1}{k_0} M^{-1} B a_0(\tau)\sinh k_0(\sigma-\ell/2)] \\ &\quad + \sqrt{\frac{2}{\ell}}\sum_{n=1}^\infty [c_n(\tau)\cos\frac{n\pi}{\ell}\sigma + \frac{\ell}{n\pi} m^2 B a_n(\tau)\sin\frac{n\pi}{\ell}\sigma],\end{aligned} \tag{4}$$

in which $k_0^2 = m^2 \tilde{B}^2/(1+\tilde{B}^2)$, and $M_{ij} = \delta_{ij} - B_{ij}^2$.

Note that, the B-dependence of the above solutions is resulted just from the above BC; the B-field has no role in the equations of motion. Note also that, for $\tilde{B} \to 0$, the above mixed BC, turns into a simple Neumann BC, while for $\tilde{B} \to \infty$, it reduces to a simple Dirichlet BC, so for both limits, the dependence on the B-field, would disappear from the solutions.

Inserting the above solutions into eq. (4) gives Hamiltonian in terms of physical modes as [21]

$$H = \sum_{n=0}^\infty [C_n^2 + \omega_n^2 A_n^2] \tag{5}$$

in which

$$\begin{aligned}\omega_0^2 &= m^2(1+\tilde{B}^2), \\ \omega_n^2 &= m^2 + n^2\pi^2/\ell^2 \quad ; \quad n \geq 1\end{aligned} \tag{6}$$

and

$$\begin{aligned}A_0^2 &= \frac{(1+\tilde{B}^2)\sinh\ell k_0}{\ell k_0} a_0^2, \\ C_0^2 &= \frac{\sinh\ell k_0}{\ell k_0(1+\tilde{B}^2)} c_0^2 \\ A_n^2 &= (1+\frac{\ell^2 k_0^2}{n^2\pi^2})(1+\tilde{B}^2)\, a_n^2, \\ C_n^2 &= \frac{1+\ell^2 k_0^2/n^2\pi^2}{1+\tilde{B}^2} c_n^2 \quad ; \quad n \geq 1\end{aligned} \tag{7}$$

Hamiltonian (5) is obviously a superposition of infinite number of independent harmonic oscillators with $A$'s as positions and $C$'s as momenta, and with mode frequencies given in eq. (6).

Now, we can use these results to study the Casimir effect for the current problem. As we know the vacuum energy of a quantum field is just the summation over the zero-point energies of quantum harmonic oscillatory modes of the field. So from eq. (6), the vacuum energy of the string would be

$$\begin{aligned}E_{\text{vac}} &= \frac{1}{2}\sum_{i=1,2}\sum_{n=0}^\infty \omega_n \\ &= m\sqrt{1+\tilde{B}^2} + \sum_{n=1}^\infty \sqrt{m^2 + (\frac{n\pi}{\ell})^2},\end{aligned} \tag{8}$$

where the Planck units have been used. Note that this vacuum energy corresponds actually to two transverse directions where the coordinates $X^i$ are coupled to the B-field. Supposing that all coordinates of the 24 transverse directions of a Bosonic string are coupled to the B-field, then the vacuum energy of the string is obtained as

$$\begin{aligned}E_{\text{vac}} &= \frac{1}{2}\sum_{i=1}^{24}\sum_{n=0}^\infty \omega_n \\ &= 12\left(m\sqrt{1+\tilde{B}^2} + \sum_{n=1}^\infty \sqrt{m^2 + (\frac{n\pi}{\ell})^2}\right).\end{aligned} \tag{9}$$

The series in eq. (9) is obviously infinite, as usual in quantum field theory in assigning the ground state energy of a system. In order to regularize this series, a generalized form of the known Abel-Plana formula [3] can be used as follow

$$\begin{aligned}\sum_{n=0}^\infty G_A(n) - \int_0^\infty dt\, G_A(t) = \\ \frac{1}{2} G_A(0) - 2\int_A^\infty \frac{dt}{e^{2\pi t}-1}(t^2 - A^2)^{\frac{1}{2}},\end{aligned} \tag{10}$$

where "$k$" is a continuous variable corresponding to "$n$". To find the convergent part of the divergent series in eq. (9) we just need to take $G_m(n) = \sqrt{m^2 + (n\pi/\ell)^2}$. After some simplifications we find

$$\sum_{n=1}^{\infty} \sqrt{m^2 + (\frac{n\pi}{\ell})^2} = \frac{\ell}{\pi}\int_0^{\infty} dk \sqrt{m^2 + k^2} - \frac{m}{2} - \frac{1}{2\pi\ell}\int_{\mu}^{\infty} \frac{dy}{e^y - 1}\sqrt{y^2 - \mu^2}, \quad (11)$$

in which $\mu \equiv 2m\ell$. So considering eq. (9), the vacuum energy of the string is obtained as

$$E_{\text{vac}} = \frac{12\ell}{\pi}\int_0^{\infty} dk\sqrt{m^2 + k^2} + 12m\sqrt{1+\tilde{B}^2} - 6m - \frac{6}{\pi\ell}\int_{\mu}^{\infty} \frac{dy}{e^y - 1}\sqrt{y^2 - \mu^2}. \quad (12)$$

The first term in the right-hand side of the above equation, that is, the divergent integral, is from the vacuum energy of free space. In fact, the contribution of the free space would be obtained by turning the discrete index "$n$" into a continuous variable, and we find

$$E_{\text{vac}}^{\text{free}} = \frac{12\ell}{\pi}\int_0^{\infty} dk\sqrt{m^2 + k^2}. \quad (13)$$

Then, the Casimir energy would be obtained by subtracting the contribution of the free space eq. (13), from the vacuum energy eq. (14);

$$E_{\text{cas}}(m, \tilde{B}) = E_{\text{vac}} - E_{\text{vac}}^{\text{free}} = (12\sqrt{1+\tilde{B}^2} - 6)m - \frac{6}{\pi\ell}\int_{\mu}^{\infty} \frac{dy}{e^y - 1}\sqrt{y^2 - \mu^2} \quad (14)$$

Since actually the parameter $\mu = 2m\ell$ is very small, so the integral in the above equation, could be approximated as

$$\int_{\mu}^{\infty} \frac{dy}{e^y - 1}\sqrt{y^2 - \mu^2} \approx \int_{\mu}^{\infty} dy \frac{y}{e^y - 1} = \frac{\pi^2}{6} - \mu + O(\mu^2). \quad (15)$$

So the Casimir energy (14) takes the form

$$E_{\text{cas}}(m, \tilde{B}) \approx -\frac{\pi}{\ell} + \Delta_m \quad (16)$$

with $\Delta_m \equiv (12\sqrt{1+\tilde{B}^2} + 12/\pi - 6)m$.

Note that $\Delta_m > 0$, so the B-dependent correction term always increases the ground state energy of the string, as physically expected. For small B-field ($\tilde{B} \ll 1$) we have $\Delta_m \cong (6\tilde{B}^2 + 12/\pi + 6)m$, while for large B-field, ($\tilde{B} \gg 1$) we find $\Delta_m \cong 12\tilde{B}m > 0$. Thus for $\tilde{B} > \pi/12m\ell \gg 1$ we find a positive value for Casimir energy. This is an important result for the ground-state mass of the string. Corresponding results for the zero mass-parameter string can be obtained simply, by taking the limit $m \to 0$ in eq. (16). Note that the Casimir force (which can be defined as the derivative of the above Casimir energy, with respect to the length-parameter $\ell$), has no physical interpretation, because $\ell$ is a fundamental constant. As is seen from eq. (16), for the string with zero mass-parameter, the Casimir energy has no dependence on the B-field, however, for the nonzero mass-parameter case, the background B-field has a role in Casimir energy, so, the ground-state mass of the string has a dependence on B-field.

## 3. Casimir correction to the ground-state mass of the string

It is well known that the ground state mass of the string is given by the total zero-point energy of the string (transverse) oscillations. For our chosen parameterization for the string Lagrangian, the ground-state mass of the open Bosonic string, can be written as

$$\pi\sqrt{\alpha}M_0^2 = E_0^{\text{tot}} \equiv \sum_{I=1}^{24}\sum_n \frac{\omega_n}{2}, \quad (17)$$

In which, $\alpha \equiv \ell^2$ is the Regge slope parameter. But, as we previously mentioned, the finite value of the total zero-point energy is just the Casimir energy, so eq. (17) can be written as

$$\pi\ell M_0^2 = E_{\text{cas}}(m, \tilde{B}) \quad (18)$$

Obviously, for $m = 0$ the above equation reduces to the well-known Tachyonic (imaginary) mass of the open Bosonic string $\alpha M_0^2 = -1$. Hence using eq. (16) we can write the ground-state mass of the open Bosonic string, as

$$\pi\ell M_0^2 = E_{\text{Cas}}(m, \tilde{B}) = -\frac{\pi}{\ell} + \Delta_m \quad (19)$$

With $\Delta_m$ given by eq. (16). But, as we found before, for $\tilde{B} > \pi/12m\ell$, the Casimir energy would turn positive, thus we realize an important result: for sufficiently large B-field such that $\tilde{B} > \pi/12m\ell$, the ground state of the string contains real (i.e. non-Tachyonic) particle, with mass given by

$$\pi\ell M_0^2 \cong -\frac{\pi}{\ell} + 12m\tilde{B} > 0,$$

for (20)

$\tilde{B} > \pi/12m\ell \gg 1$.

So, the background B-field has physical importance for the ground-state mass of the string.

## 4. Concluding remarks

We have found the Casimir energy of an open Bosonic string with back-ground field, by using standard methods given in the literatures on Casimir effect, however, some results of this work are new:

We have interpreted the Casimir energy as the total ground-state mass of the string. Consequently we have obtained new *B*-dependent correction terms for the mass of Bosonic string, which are important specifically for large B-field. In fact, we have shown that, for $\tilde{B} > \pi/12m\ell$, the Casimir energy would attain positive value, i.e. for sufficiently large B-field, the ground state of the string contains real (i.e. non-Tachyonic) particles. Therefore the background B-field has physical importance for the ground-state mass of the string. It is interesting to note that, the background B-field has been introduced just through the boundary conditions, so the $B$-dependence of the ground-state mass is just due to the specific boundary condition imposed on the string.